\documentclass[onecolumn]{aastex63}

\usepackage{xcolor}
\usepackage{color}
\usepackage{ulem,xpatch}
\usepackage{float}
\usepackage{tablefootnote}

\def\eg{{\it e.g.,} }

\def\ie{{\it i.e.,} }

\def\heii{He\,{\sc ii}}

\def\nii{N\,{\sc ii}}

\def\sii{S\,{\sc ii}}
\def\siii{S\,{\sc iii}}

\def\oi{O\,{\sc i}}
\def\oii{O\,{\sc ii}}
\def\oiii{O\,{\sc iii}}

\def\ariv{Ar\,{\sc iv}}
\def\cliii{Cl\,{\sc iii}}

\def\kms{km\hspace{1pt}s$^{-1}$}
\def\cc{cm$^{-3}$}

\usepackage{color}
\def\lsim{\lower.5ex\hbox{$\; \buildrel < \over \sim \;$}}
\def\gsim{\lower.5ex\hbox{$\; \buildrel > \over \sim \;$}}

\definecolor{violet}{rgb}{0.53, 0.00, 0.69}

\received{October 10, 2022}
\revised{December 19, 2022}
\accepted{December 19, 2022}

\submitjournal{ApJS}

\begin{document}

\title{The [{\sii}] Spectral Images of the Planetary Nebula NGC 7009: {{\large\sc ii}.} Major Axis 
}

\def\corrauthor{
S. Hyung
}


\def\runningauthor{
Hyung et al.
}


\def\runningtitle{
[{\sii}] Images of NGC 7009 Major Axis 
}

\correspondingauthor{Siek Hyung}
\email{hyung@chungbuk.ac.kr}

\author{Siek Hyung}
\author{Seong-Jae Lee}
\affiliation{Dept. of Earth Science (Astronomy),
  Chungbuk National University, Chungbuk 28644, S. Korea}

\author{Masaaki Otsuka}
\affiliation{Okayama Observatory, Kyoto University, Kamogata, Asakuchi, Okayama, 719-0232, Japan}

\begin{abstract}
We derived position-velocity density distribution diagrams along the major (PA = 77$^{\circ}$) axis of the elliptical planetary nebula NGC 7009 with the Keck HIRES [{\sii}] 6716/6731\AA\, doublet spectral images.  
The average densities of the main shell and knots of NGC 7009 derived from the [{\sii}] 6716/6731\AA\, fluxes integrated over the line of sight indicate a density range of  $N_{\rm e}$ = $10^{3.4}$ to 10$^{3.9}$ {\cc}, while the local densities from the volume fraction resolved in radial velocities along the line of sight show a considerably large range of about 10$^{2.8}$ -- 10$^{4.7}$ {\cc}. The derived projection angle of the major axis of the main shell is about $\psi  \sim$18.3($\pm$2)$^{\circ}$. 
Assuming that the main shell is an ellipsoidal shell with $a 
\simeq$16$''$ and $b \simeq 6''$, we found the range of expansion velocity, radius, and latitude of four knots and a hot bubble.
The four knots at the points in symmetrical positions are roughly aligned with the same axis of expansion of latitudes $\phi \sim \pm 34.5(\pm 0.6)^{\circ}$: One pair expands at about 35 {\kms} close to the main ellipsoidal shell, and the other expands rapidly at about 60 {\kms} at a distance of $r \sim 16''$.
In the latitude range $\phi = $65 -- 75$^{\circ}$, the hot bubble of a relatively large structure expands rapidly with a velocity of 130 -- 150 {\kms}.
Four knots and hot bubble points that expand faster than the main shell appear to have been formed by two to three eruptions at a different epoch than the primary structure formation.

\end{abstract}

 
\section{Introduction}

The Saturn planetary nebula (PN) NGC 7009 has a relatively large angular size sufficient to investigate details of the substructure associated with kinematic and evolutionary status. The morphological shape is generally classified as either an early intermediate elliptical or an intermediate ellipse, with fast isotropic stellar wind on the inner part and shells surrounded by an external shell  (\citealt{1987B1}).
Observational data on bright rims, knots, and halos obtained with high-dispersion long-slit spectrometers,  Fabry-Pero interferometry, or narrow filters, allow better study of kinematic structures of the main shell and outer halo regions  (\citealt{1985RA, 1994boh, lame1996}).

\citet{Gon03,Gon04} identified several small substructures of inner and outer knots (caps and ansae) and jetlike (filamentary) infrastructures besides the bright circular rims from the 1996 $HST$ morphological images with the [{\oiii}] and [{\nii}] narrow-band filters.  \citet{2002Gue} detected the diffuse X-ray emission of 1.8 $\times 10^6$ K and the hot gas within the central cavity. 
\citet{ri13, ri14} derived position-velocity (PV) density diagrams of 
electron temperature and density along the major axis with the ESO Very Large Telescope UVES slits.  
\citet{sab04} presented a three-dimensional (3D) density distribution model structure based on the kinematic result of high-dispersion echelle spectra.

\citet{1985RA} derived the expansion velocities of the inner and outer knots of $\pm$60 {\kms} in the Fabry-Pero [{\oi}] 6300\AA\, line, based on the earlier proper motion measurements of \citet{1965liller}, assuming an inclination angle $i$ = 50$^{\circ}$. 
\citet{2022Lee} found the expansion velocities of 19.19 -- 21.70 and 
24.34 -- 30.00 {\kms} along the minor axis for the main and outer shells, respectively. 
The acceleration of the inner main shell is expected in the usual interacting winds scenario \citep{1978kwok}. 

Considering the axial ratio of 2.5 between the major axis and the minor axis of the main shell ellipsoid, the maximum expansion velocity of the main shell along the major axis is 54 {\kms} at least 2.5 times that of the minor axis.
\citet{ste09} estimated the maximum velocity of the main shell near the polar apex as 60 -- 90 {\kms} based on the observations of expansion velocities and proper motions (\citealt{sab04, 2004Fern}). Any error in our assumption about the expansion velocities of the primary apex would not change our conclusions about the kinematic properties of knots and hot bubbles.

Most recently, Lee et al. (2022, Paper I) found that local densities of the minor axis of NGC 7009 decomposed in velocities along the line of sight, whose range is very wide compared to the mean density derived from the integrated flux for the line of sight.
We will derive the PV density distribution diagrams along the major axis. Similar to Paper I, 
we will analyze the two-dimensional (2-D) [{\sii}] 6716.4 and 6730.8\AA\, spectral images of the major axis of NGC 7009, secured with the HIRES (HIgh-Resolution Echelle Spectrograph), attached to the Keck 10-m telescope.  
We assume that the main shell of NGC 7009 is a simple prolate structure and determine the projection angle of the main shell.
We will also study the kinematic characteristics of knots and a hot bubble observed in the major axis.

\section{Keck HIRES observations}

The major axis of NGC 7009 was observed with Keck HIRES in 1998 August 15 (UT)  by Hyung and  Aller. The seeing was slightly less than 1.0$''$.
Since the bright nebula rim size along the major axis is about 28$''$ $\times$ 30$''$, three arrangements of the 14$''$ $\times$ 0.862$''$  slit were placed to cover the whole major axis. 
Figure 1 shows an NGC 7009 HST image and the positions of the Keck HIRES slit on the major axis. 
The position angle (PA) (or the slit direction) superimposed on the nebula image is PA $= +77^{\circ}$ and the slit width 0.862$''$ corresponds to the wavelength dispersion power, 
$R = \lambda$/$\Delta\lambda$ = 45\,200 ($\sim$3 pixels on a CCD). 
Table~\ref{tbl-1} lists which part of the PN was observed by the Keck HIRES slit. 
Column (2) shows the distance from the CSPN (central star of PN) to the center of the 14$''$ HIRES slit (either W-SW or E-NE), and column (3) shows the range the analyzed final spectral image.
For detailed slit locations and observed areas with slit entrances, see also Figure~\ref{fig2}.

\begin{figure*}[h]
\centering
\includegraphics[width=0.6\textwidth]{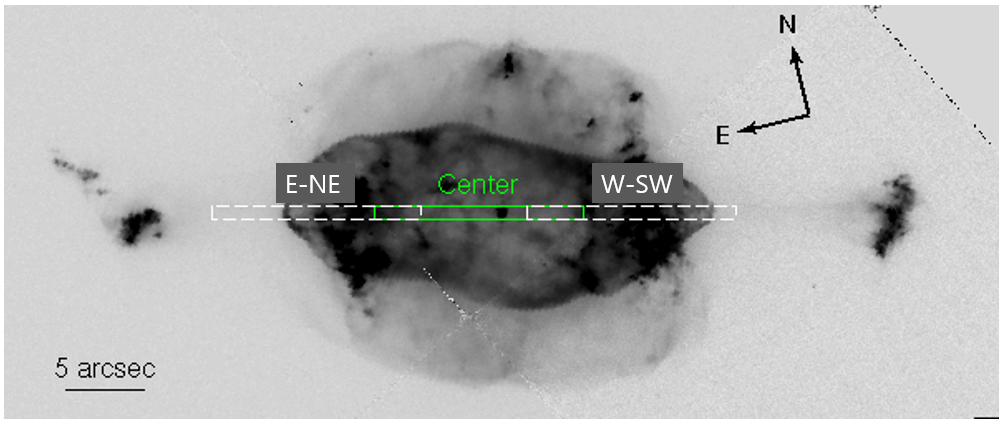}
\caption{HST image of NGC 7009 and Keck HIRES slit direction on the nebular image. The HST WFPC2 image with the F658N filter shows main shell, outer shell, and caps/ansae  (NASA/ESO).  The 14$'' \times$ 0.862$''$ slit entrance was placed at the W-SW, the center (Cen),  and the E-NE areas along the major axis  (PA $= +77^{\circ}$). See Table 1.
 }
\label{fig1}
\end{figure*}

\begin{table}[ht]

\caption{Keck HIRES slit setup.}
\begin{center}
\begin{tabular}{@{}lcccc@{}}
\hline
{Slit} &     slit center   & spectrum (pixel) & note \\
\hline
W-SW    &      +8.7$''$ (W-SW)  &   +2.4$''$(2) -- +15.3$''$(1) & (a), (d) \\
Center  &      $-$1.5$''$ (E-NE)     &  +4.8$''$(0) -- $-$8.4$''$(2) & (b), (e) \\
E-NE    &      $-$12.3$''$ (E-NE) &  $-$5.3$''$(1) -- $-$15.7($''$10) 
 & (c), (f) \\
\hline
\end{tabular}
\label{tbl-1} 
\end{center}

Notes: Column (2): The distance  ($''$) to the slit center from the central star; \eg +8.7$''$(W-SW) means the slit center in Figure 2(a) is along the major axis towards 8,7$''$ W-SW of the star.
Column (3): y$''$ range of the spectral image (and discarded pixels near the boundary) in Figures 2(d), (e), and (f). For example, +2.4$''$(2) and +15.3$''$(1) indicate that the spectral image of Figure 2(d) in a range from +2.4$''$ to +15.3$''$ relative to the central star,  obtained after discarding 2 pixels and 1 pixel near the upper and lower boundaries, respectively (1 pixel $\simeq = 0.37''$). 
See Figure~\ref{fig2} slit dimensions placed along long axis PA = +77$^{\circ}$.
Column (4): Figure 2 for columns (2) and (3).
\end{table}

Figures~\ref{fig2}(a), (b), and (c) show two images:   the left panel shows the yellow-green-blue false color representation (weak flux to strong flux), with the slit entrances placed at the  W-SW apex, CSPN, and  E-NE apex, respectively, over the PN image taken during the actual observations.
Figure 2 of Paper I also presented slit monitoring images similarly. 
Those images recorded on the Keck HIRES slit monitor are reversed left and right as if they were displayed in a mirror. In Paper I, the authors presented the slit position images without correcting the reversed left and right directions. On the other hand, in Figure 2 of this paper, the mirrored images recorded on the slit monitor switched back to normal to match the HST image in Figure 1.
False orange indicates sky background areas with no signal.
Because the nebular image with the slit entrance was taken with the camera positioned slightly sideways, the slit entrance does not appear exactly along the center of the major axis. 
The right white-blue false colors in (y-axis) 14$''$  $\times$ (x-axis) 3\AA\ panels are the reduced  [{\sii}] 6731\AA\, images.
The two dotted arrows in the upper left panel show the portion (spatial length $\times$ width) of the surface of the planetary nebula observed through the slit entrance corresponding to the upper right resulting spectral image. 
When finding a PV density counter diagram, we discarded 1 -- 2 pixel(s) in case the vicinity of both boundaries was chaotic. 
All three observations were  2-minute exposures.  
Paper I shows that the 2-minute exposures give a fairly good signal-to-noise ratio compared to the other 25-minute exposures.  Looking at the spectral images on the right of Figures 2(a), (b), and (c), the spatial resolution of the spectral image in (c) is relatively poorer than the other two. This difference is likely due to the seeing condition deteriorating at this time. 
The reduction procedure and other information are given in Paper I.

Similar spectral images along the major axis can be found in previous works, \eg \citet{sab04} and \citet{ri14}.
This study updates previous observational studies based on [{\sii}] spectral images.
The bright primary shell shapes are roughly perceived as ellipsoids, \eg in \citet{sab06} and \citet{ste09}.
The spectral images at the end of the major axis in Figures~\ref{fig2}(a) and (c) show two bright knots, namely the blue side W-SW and the red side E-NE blobs, which expand radially outward symmetrically with respect to the CSPN.

\section{PV density distribution diagrams}

The present work extends the Keck HIRES data analysis of the minor axis data by Paper I. 
Hyung and Aller (1995a) derived $T_{\rm e}$({[\nii}]) $\sim$ 9000 -- 10\,300~K for the W-SW bright position of the bright rim apex along the major axis. Meanwhile, \citet{ri13} derived 
$T_{\rm e}$ $\sim$ 8500 -- 10\,500~K from [{\nii}] and [{\oiii}] diagrams (see their Figures 5 and 6). 
\citet{2022Akr} derived 
an almost constant temperature $T_{\rm e}$([{\siii}]) of   about 9200K for distances up to 22$''$ for a the slit at PA = 79$^{\circ}$. $T_{\rm e}$({[\nii}]) also shows a steady fluctuation of $\sim$10\,000~K in the range $r =$ 5 -- 15$''$ but has a chaotic change of $> $11\,000 K in the inner central zone and the outer zone (see their Figure 6). 
The HES data observed at the Lick observatory by \citet{ha95a} reported mean densities of log~$N_{\rm e}$  = 3.4 dex ([{\ariv}]), 3.65 ([{\cliii}]), 3.63 ([{\oii}]), 3.75 [{\sii}]) within the 1.2$'' \times 4''$ slit entrance at the W-SW  apex of the major axis.

The mean density derived from the [{\sii}] 6716/6731 flux integrated at each slit location along the minor axis (PA = 347$^{\circ}$) is log~$N_{\rm e}$ = 3.7 -- {4.1} dex (see Figure 8 in Paper  I).
\citet{2022Akr} induced a density range of log~$N_{\rm e}$ = 3.5($\pm$0.02) -- 3.8($\pm$0.07) dex from the integrated [{\sii}] lines along the major axis (PA = 79$^{\circ}$) (see their Figure 6).
Adopting $T_{\rm e}$({\nii}) $\sim$ 9500~K, we found similar densities of log~$N_{\rm e}$ = 3.4 -- 3.9({$\pm0.1$}) dex along the major axis (PA = 77$^{\circ}$) for Figure 2(e).

As explained in Paper I, Keenan et al. (1996, K96) showed that the  [{\sii}] 6716/6731 ratio from 3p$^{3}$ $^{2}$D{\tiny 3/2,5/2} to the ground state  3p$^{3}$  $^{4}$S{\tiny 3/2} involves the other high metastable transition, 3p$^{3}$  $^{2}$P{\tiny 1/2, 3/2}. As a result, the nebular diagnostic  [{\sii}] 6716/6731 ratio can give the density information up to log~$N_{\rm e}$  =  {5.0} dex (for an assumed electron temperature). 
The process until the PV density diagram came out had been fully discussed in Paper I. 
We obtained the density diagram using Cloudy, C17.02, which reflects the research results by \citet{1996Keenan}. The derived density by 
Cloudy, C17.02,  was confirmed to be the same as the value obtained using PyNeb. 
Like Cloudy, \citet{2020Mori}'s PyNeb reflects the high-level aurora-to-nebula transition's contribution to the [{\sii}] nebula transition introduced by K96.
As a result, the density up to log~$N_{\rm e}$ = {5.0} dex can be obtained only with [{\sii}] 6716/6731 nebular line ratio.  

C17.02 uses collisional strengths from \citet{2010TZ}.
Meanwhile, C17.02 uses [{\sii}] transition rates from \citet{2005iri} instead of \citet{2014Ki} because the former is believed to be optimized for forbidden lines.     
\citet{2020Mori} conducted a study comparing [{\sii}]6716/6731 emissivities predicted by the transition rates obtained from various studies (see their Figure 6 left panel and Table 2).
In the density region above log~$N_{\rm e}$ = {4.2} dex, the transition rate by \citet{2014Ki} gives approximately 24\% higher emissivity than that by \citet{1982MN}, while the transition rates by \citet{2005iri} match within a few \%.
However, in the density range of log~$N_{\rm e}$ = 2 -- 4 dex, the transition rate of \citet{2005iri} 
is 5 -- 8\% lower than predicted by \citet{1982MN} or most others.

As mentioned, K96 calculated the collisional strength using the excitation rates of \citet{1996Ram} (see K96's Table 1). 
If we compare the emissivity calculated by K96 with the results presented in \citet{2020Mori}'s Figure 6, we can see that it resembles \citet{2019RG} or \citet{1993KH} and, in particular, agrees very well with \citet{2019RG}. 
Using the atomic constants of K96 or \citet{2019RG}, the derived density would become higher by 0.1 -- 0.2
dex than the present value (see Figure 3 in Paper I). 
In short, the actual density might be at least 0.1 dex higher depending on the choice of atomic constant, but this is unlikely to change the conclusion.

As mentioned, we used the Cloudy photo-ionization (P-I) modeling to find the electron densities up to log~$N_{\rm e}$  =  {5.0} dex [{\cc}] from the [{\sii}] 6716/6731 ratio. 
However, if one employed PyNeb, one could derive the line emissivities and density information like ours more efficiently.
Assuming the $T_{\rm e}$ $\sim$ 10\,000~K, Paper I derived the local densities of the equatorial zone of the minor axis by
[{\sii}] line ratios are log~$N_{\rm e}$  =  2.8 -- 4.7 dex {\cc} resolved in radial velocities along the line of sight, much larger than the mean value suggested by the Lick observatory HES  log~$N_{\rm e}$  =  3.8$\pm 0.2$ dex {\cc} or the above mentioned mean density from the integrated fluxes. 

\begin{figure*} 
\centering
\includegraphics[width=185mm]{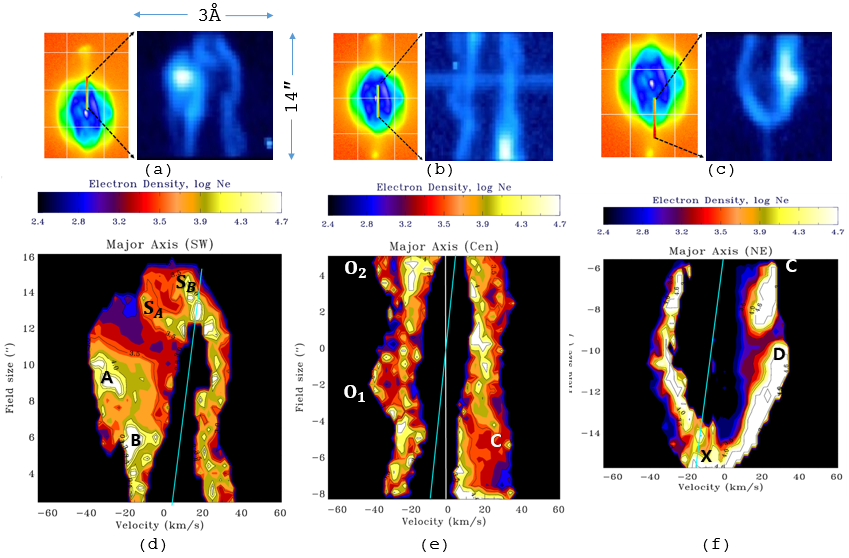}
\caption{
Keck HIRES [{\sii}] 6731\AA\, 2-D spectra, and PV density diagram along the major axis (PA =  +77$^{\circ}$; 1998 August 15 UT). All the used slit lengths are 14$''$ long, and all the exposures are 2 minutes. (a), (b),  and (c):  
the left panels show the nebular image with the 14$'' \times 0.862''$ slit entrance placed along the major axis and  
the right panels show the close-up [{\sii}] 6731 2-D spectral flux images in logarithmic scale from the left panel, 
obtained from the slit entrance placed at the W-SW, CSPN, and E-NE positions along the major axis. 
West (W-SW) is upward and East (E-NE) is downward; North (N-NW) is left and South (S-SE) is right. 
(d), (e), and (f): PV density diagrams derived from the [{\sii}] 6716/6731\AA\, ratio. 
The white vertical line in (e) indicates V$_{\rm r}$ = 0.0 {\kms}.   The tilted sky blue line is a redefined kinematic base line of the prolalte ellipse. 
See the text.
 }
\label{fig2}
\end{figure*}

Similarly, we can derive the local density of NGC 7009 from each volume slice (resolved by radial velocities) along the line of sight at selected slit locations. 
Figures~\ref{fig2}(d), (e), and (f) show the derived PV density distribution diagrams where the horizontal x-axis represents the radial velocity along the line of sight. 
The horizontal line-of-sight radial velocity corresponds to the apparent expansion velocity in {\kms} relative to the kinematic center (of the moving frame), while the vertical y-axis corresponds to the spatial dimension along the major axis direction relative to the CSPN, \ie (+) for the W-SW and ($-$) for the E-NE. 
We added a vertical line (white) to the center of Figure~\ref{fig2}(e), which shows a temporary correction using a radial velocity of $-$48.91 {\kms} (see Paper I).
The false colors represent the density values.

We set the lower limit of the  [{\sii}] 6716 and 6731 by 1.5\%, 2\%, and 1.5\% of the flux maxima in deriving the  PV density  diagrams in Figures~\ref{fig2}(d), (e), and (f) to avoid confusing boundaries due to extremely low S/N values. 
The white color zones indicate the regions of highest density as high as log~$N_{\rm e}$ = {4.7} dex (50\,000) {\cc}, while the blue color ones imply the low density regions log~$N_{\rm e}$ = {2.8} dex {\cc}. The estimated errors in the derived densities are about 0.03 dex.

Our [{\sii}] 2-D spectral images  along the PA = 77$^{\circ}$ correspond to [{\nii}] image at PA = 79$^{\circ}$ by \citet{sab04} (see their Figure~2).
The 2-D spectral images show an elliptical shape of the primary shell
seen in other [{\oiii}]  and {\heii} images by  \citet{ri13, ri14}.

We are able to see the outer shell near the equator in Figure~\ref{fig2}(b). 
Figure~\ref{fig2}(e) clearly shows a semi-circular structure at the equator and the detached sub-shell structures at low latitudes, \ie blue shifted O$_1$ and O$_2$ tori expanding with -37 {\kms} at $-2''$ (O$_1$) and -42 {\kms} at 4.5$''$ (O$_2$) at low latitudes. The latter tori expand faster than the outer shell at 30 {\kms}, identified in Paper I. 
Due to incomplete stellar continuum subtraction, the outer equatorial shell appears weakly on the blue-shift side and does not separate from the main shell on the red-shift side. In fact, the outer shell at the equator is not a perfect circle but is made up of broken substructures.  The presently identified two tori expand faster than the equatorial outer shell (\citealt{ste09, sab04}).
The  [{\sii}] PV density diagram in Figure~\ref{fig2}(e) implies that the bright main and faint outer shells consist of some small zones in the density range of log~$N_{\rm e}$ = 2.8 --  4.6 dex  {\cc}.

Pairs of low-ionization knot, often called “FLIERs” (fast, low ionization emission regions), are seen in their [{\nii}] and [{\oiii}] line images, along the major axis of fully ionized PNe \citep{1994Bal}.
In Figure~\ref{fig2}, we marked `A,'  `B,'  `C,' and `D,' for four salient substructures of the notably different kinematics. 
Knots 'A' and 'D' are well-known FLIERS. We also identified other 'B' and 'C' structures as knots with kinematic properties similar to FLIERS. These four structures are not found in the cap location. Our observations show that a hot bubble structure appears in the cap. 
The location of `C' is marked in Figures 2(e) and (f). As mentioned above, the PV density contour of `C' of a relatively weak intensity does not appear in Figure 2(f) due to the poor spatial resolution. Therefore, we used Figure 2(e) for the kinematic information of `C.' 
Figure 2 shows that the blue shift two knots `A' and `B' have dimensions of about 18 -- 20 {\kms} $\times$ 2$''$ and 10 -- 12 {\kms} $\times$ 1.5$''$ respectively. 
Meanwhile, the other two less prominent red shift knots `C' and `D' are about { 14 -- 15 } {\kms} $\times$ 1.5$''$ and {10 -- 12 } {\kms} $\times$ 1.5$''$  respectively.  
Two pairs of knots `A' and `D' and knots `B' and `C' at point symmetric positions, appear to expand radially outward from the center. 
These knots could be 
explained as the very dense, cold, and neutral remnants of
magnetically formed knots (\citealp{2020B, Gon03}).

`A' and `D'  at the W-SW and E-NE sides, are the previously identified well-known FLIERs (knots), \eg [{\nii}] and [{\oiii}]   (\citealt{sab04, sab06, ri13}).
The different expansion velocities and spectral images in Figures~\ref{fig2}(d) and (f) imply that these knots belong to neither the primary shell nor the faint outer shell. 
The knot `D'  must be the counterpart substructure of the knot `A'. 
Two fast `A' and `D' knots appear to be outside the bright rim.     
The pairs of knots and ansae along the major axis are perhaps due to the mass-loaded stellar winds \citep{1985RA, 1987B2BPV}.
`X' is marked for a specific region in Figure 2(f), at the counterpart position S$_{\rm A}$ and S$_{\rm B}$.
The peculiar features of this area are not clearly visible in the spectral image in Figure 2(c), but 
seem to show up in the PV density diagram. Perhaps the stellar winds do not yet play a role in developing a hot bubble structure.

Figure~\ref{fig2} shows overall elliptical expansion kinematics relative to the newly determined kinematic central axis or the sky blue tilted baseline. The main shell shown in Figure 2 is too complex to be depicted as something as simple as an elliptical geometry. To study the complex structure shown in Figure 2 in detail, we first assume that the gas emitted from the central core expands to have an elliptical structure to form the main shell shown in the 2-D PV diagram (see also \citealt{ste09}). 
We will then proceed with a detailed analysis of the rate of expansion with latitude based on the geometry of a prolate ellipsoid or sphere.

For the emitted gas of the central star to evolve into the prolate ellipsoidal shell, the differential expansion velocity at a distance from the center must be closely related to the spherical shell coordinates ($r$, $\phi$): the shape of the prolate shell is formed when the expansion speed of each point of the shell is proportional to the radius ($r$) from the center. 
When the major axis of the prolate spheroid shell is projected parallel to the two-dimensional sky plane, this prolate spheroid will appear as an ellipsoid with the major axis ($a$) and minor axis ($b$) undistorted. In this case, each point of the ellipse can be expressed in a Cartesian coordinate system ($x$, $y$) or a spherical coordinate system ($r, \theta$).
Paper I shows that the expansion velocity   21.7 {\kms} for ($r, \theta$) = (6$'', 0^{\circ}$), i.e. V$_{\rm x} $ at the equator = 21.7 {\kms} and V$_{\rm y}$ = 0 {\kms} (see Appendix A).

\begin{table}
\caption{ Four knots and hot bubble.  }
\begin{tabular}{@{}lcccccc@{}}
\hline 
Knot &  Distance  & log~$N_{\rm e}$ & {$\theta$ } & EM1 (V$_{\rm x}$,V$_{\rm y}$) & V$_{\rm x}$(obs)  & $|\Delta$V$|$    \\
  (Figure 2)   &($''$)      &  [{\cc}] &  ($^{\circ}$)  &    &   &   \\
\hline
A  &  8.5 -- 10 [9.3] &  3.8(0.514) -- 4.7(0.443)  &   63.2  & (-17.0, 33.6) & -37$\pm$6  &     20 \\
B  &  5 -- 6 [5.5] &   3.8(0.514) -- 4.7(0.443) &  44.6   &  (-20.2, 19.9) & -18$\pm$3  & 2.2 \\
C  & -5 -- -6 [-5.5] &  3.5(0.584) -- 3.9(0.498) & -44.6   & (20.2, -19.9)   &    ~25$\pm$5   & 4.8   \\
D  & -9.5 -- -11 [-10.3]  &  3.5(0.584) -- 4.7(0.443) & -{67.0}   & (15.8,  -37.3)  &   ~35$\pm$3  & 19.2  \\
S$_{\rm A}$(S$_{\rm B}$)   & 13  &      3.5(0.584) -- 3.9     &    $77.0$ & (-10.8, 47.0) &   -14,+5($\pm$3)  & 3.2(-15.8) \\
\hline
\end{tabular} 
\label{tbl-2} \\
\vspace{0.05in}

{Notes:  All Velocity units: {\kms}. See Appendix A.
The projection of the elliptical shell ($a$ = 15$''$ and $b$ = 6$''$) onto the sky plane is not considered, here.
Column (1): indicator letters are given in Figure 2. 
Column (2): the vertical distance from the center  (+ for W-SW; and $-$ for E-NE) with [the median value]. The positions of four knots and a hot bubble structure are in Figure 2.
Column (3): logarithmic densities: 3.8(0.514) means log $N_{\rm e}$ = 3.8($\pm$0.1) dex and [{\sii}]6731/6716 = 0.514. Log $N_{\rm e}$ 4.7(0.443) means the maximum density zone, and the line ratio of this area is slightly larger than 0.443, which means that its density is, in fact, in the range of 4.6 - 4.7 dex. Column (4): latitude for the median value in the elliptical model (EM1) in Appendix A. Column (5): the calculated x- and y- velocity components. The  $x$-component of EM1 corresponds to the radial velocity.
Column (6):  the observed radial velocities of the knots relative to the tilted kinematic center. Column (7): $|$V$_{\rm x}$ - V$_{\rm los}|$ }
\end{table}

Table~\ref{tbl-2} indicates the density of four knots `A' -- `D' and hot bubble trace S$_{\rm A}$ and S$_{\rm B}$, whose approximate spatial distance from the CSPN along the major axis (y) is on the second Column. 
The circular substructure seen in W-SW  PV diagrams is likely to be a shock structure, formed similarly as the X-ray emitting hot bubble \citep{2002Gue}.  We list densities for S$_{\rm A}$ and S$_{\rm B}$.

Under the assumption that the major axis is parallel to the sky plane, we obtained the expansion velocity and component of the elliptical shell using the elliptic model EM1 in Appendix A.
Since the x-axis coincides with the line of sight direction, we also can find the expansion velocity from the latitude on the EM1.
The latitudes and velocity components were presented in columns (4) and (5) of Table 2. 
The last column gives the difference between the observed radial velocity in column (6) and the horizontal velocity in column (5). 

The simple model EM1 suggests the expansion of knots `B' and `C'
resembles an ellipsoidal-shaped main shell. The rest of the knots, on the other hand, appear to expand completely differently at greater distances from the main shell.
The knots are well separated from the primary shell, but the various parts of the slowly expanding primary shell seem to be affected by the fast knot.  
A hollow knot structure containing S$_{\rm A}$ and S$_{\rm B}$ at the W-SW apex identified in other studies will be called a hot bubble.

\section{Main Shell and Five Knots}

In the previous section, we checked the case of an ellipse that did not reflect the tilt of the principal axis to the sky.
In this chapter, we will explore the geometrical structure of knots and a hot bubble using the ellipse and circle models considering the projection angle.

\subsection{Two Knots near Main Shell}
The EM2 in Appendix B schematically shows a prolate elliptical shell projected onto the sky plane.  We choose a point near the knot `B' and explore its expansion velocity. The vertex position of the EM2 close to $S_{B}$ in Figure~\ref{fig2}(d) expands at about +17 {\kms} relative to the kinematic center 0 {\kms}, while the other symmetric vertex position  close to $X$ in Figure~\ref{fig2}(f) expands at about $-$17 {\kms}.

We found that the projection angle of the major axis to the sky plane is $\psi$ = 18.3($\pm$2)$^{\circ}$ in Appendix B. So, one must consider a projected ellipsoid on the sky plane, \ie EM2, the ellipse with $a \sim$ 16$''$ and $b \sim$ 6$''$ for the prolate primary shell.

Table~\ref{tbl-3} lists the derived expansion velocities of two knots `B' and `C.' 
The latitude and velocity of the knot are determined depending on which values of $d$(ma) and $r$ are selected. In other words, the two variables $d$(ma) and $r$ act as constraints that cause the model values to the knots size and velocity range.
Comparing the results shown in Tables~\ref{tbl-2}  and Table~\ref{tbl-3}, the difference between the radial velocity component by EM2 and the observed radial velocity is greater than that of EM1. However, the EM2 appears to be a more accurate reproduction of the real world, meaning that the two knots `B' and `C' are not part of the main ellipsoidal shell.

Each knot has a range of radial velocities, $\pm \Delta$V$_{\rm los}$.  
The projected velocity along the line of sight from the radially outward expansion velocity of the $\beta$ point of the EM2 shell is $-$15.6 {\kms} for knot `B'. The observed radial velocity is  V$_{\rm rad}$(obs) = $-18\pm 3${\kms}. Similarly, we can find model predictions for knot `C'.
The two knots, `B' and `C,' with a relatively small  $|\Delta$V$_{\rm los}|$ in Column (7) look close to the prolate ellipsoid. Hence, we list the expansion velocities for `B' and `C,' in Column (8).

The radial velocity differences of two knots `A' and `D' at 
a distance of 9.8$''$ and 10.0$''$, respectively, from the CSPN along the major axis, are much larger than those for `B' and `C' at 5.8 (or 6.3)$''$ and -5.8 (or -5.3)$''$, \ie $\Delta$V$_{\rm los}$  = 28.4 -- 28.8  {\kms} for `A' and `D' and 2.4 -- 9.4 {\kms} for B' and `C' (see Tables~\ref{tbl-b} and \ref{tbl-c} in Appendix).
$|\Delta$V$_{\rm los}|$ for hot bubble points S$_{\rm A, B}$ is also large, implying this structure is not also close to the ellipsoidal main shell.

\begin{table}[ht]
\caption{Expansion velocities of two knots.
 }
\footnotesize

\hspace{-0.7in}
\begin{tabular}
{ccccc cccc}
\hline 
Knot    & $d$(ma)$''$ &  V$_{\rm rad}$(obs)   & EM2($r'', \phi_{\beta}^{\circ}$)   & (EM2)V$_{\rm exp}$ & 
(EM2)(V$_{\rm los}$, V$_{\rm sky}$)  & $|\Delta$V$_{\rm los}|$   & V$_{\rm exp}$(knot)   & Note \\
\hline
B        & 5.8  & -18$\pm$3  &    {\bf (7.0, 33.6)} & {\bf 25.3}  & (-15.6, -19.9)  &   2.4  & {\bf 29.2}    & $r_{\rm max} = 8.5$ -- $8.8''$      \\
  & 6.3    & & (7.3, 37.3)   & 26.2  & (-14.8, -21.6)  & {\bf 3.2}  & 31.9     & [29.2--31.9] \\
C       & -5.8 & 25$\pm$3  &     {\bf (7.0, -33.6)} & {\bf 25.3}  & (15.6, 19.9) &    9.4  & {\bf 40.5}      &  $r_{\rm max}$  = 8.3 -- 8.5$''$    \\
    &  -5.3  &  & (6.8, -29.8)  &  24.4  & (16.3, 18.2) &  {\bf 8.7} & 37.4   &  [37.4--40.5] \\
 \hline
\end{tabular}  \\
\vspace{0.05in}

{Notes.  All Velocity units: {\kms}. 
Knots `B'  and `C'  were analyzed with the elliptical model EM2 in  Appendix B.
Column (2): Spatial distance from the center to the projected point along the major axis, \ie $y'$ in Appendix B.
Column (3): the observed radial velocity relative to the major axis corresponds to the radial velocity along the line-of-sight, \ie V$_{\rm rad}$(obs) = V$_{\rm los}$(EM2). 
Column (4): radius and latitude angle $\phi_{\beta}$ of a point on EM2 near the knot `B'. 
Similarly, the point `c,' can be selected near the knot `C'. 
Column (5): velocity values on the EM2 shell.
Column (6):  The line of sight and sky plane velocity components by EM2 (see Figures~\ref{figB1} and \ref{figB2})
Column (7): difference of the line-of-sight velocities by the model (m) and observed knot in Figure 2. Column (8): derivation of the knot's expansion velocity (and their range), \eg  V$_{\rm exp}$(B) =  V$_{\rm los}$(B)/cos($ \phi_\beta'$).
Column (9): the maximum radius has been determined by adding the approximated size of the knot to the radius in Column (4). See Appendix B.  }
\label{tbl-3}
\end{table}

The predictions for knots `B' and `C' by choosing two representatives $d$(ma) would give ranges of 29.2 -- 31.9 {\kms} and 37.4 -- 40.5 {\kms}, to comply with the observed velocity ranges. The predicted expansion velocity of the knots is approximately twice faster as the main shell. 
Note that the dimension of C and D is about 1.5$''$ in Figure 2. Thus, we assume that the maximum distance from the center to C and D is about 1.5$''$  greater than the nearby points in the EM2 shell.
Through predictions consistent with observations, EM2 shows that the maximum radii of knots `B' and `C' are 8.5  -- 8.8$''$ and 8.3 -- 8.5$''$, respectively, which are fairly close to the ellipsoidal shell, but not part of the shell as seen Figures~2(d) and (f).

\subsection{Three Knots in a Circular Model}

Since the other two knots and the hot bubble are not close to the EM2-like main shell, another geometric approach is needed to analyze their kinematic properties. So, we consider an imaginary circle that includes the knots `A' or `D,' as in Appendix C. The CM3 diagram of Appendix C shows the direction of expansion and latitude for knots `A' or `D.' 

As in EM2, we first need to select $d$(ma) for each knot, \ie `A' or `D' according to Figure 2. With the help of an imaginary circle, 
CM3 will predict the expanding speed and its components for the selected $r$ for knot `A' or `D'. 
The observed velocity and spatial range for each knot are constraints for choosing an appropriate $r$.

Similar to the EM2 case, first consider the projected point $\alpha_1$ of knot `A' on the major axis, and then various variables are obtained sequentially through the relationship with other surrounding points.
We then derived the model-predicted line of sight velocities and components calculated under the prerequisites of the observed radial velocities. 
Choosing a smaller $r$ predicts a higher expansion velocity.
After trying multiple cases, one will find a reasonable radius and velocity range suitable for the observed values.

Table~\ref{tbl-4} shows that knot `A' expands at a  speed of 55.1  -- 72.5 {\kms} at a radius of 15 -- 16$''$, while the other apparent counter position knot `D' expands relatively slowly at a speed of 57.8 -- 68.8 {\kms} at a radius of 16.5 -- 17.5$''$. 
Looking at the kinematic characteristics, knots `A' and `D' and knots `B' and `C' appear to form a pair, respectively, and the former is more active at a large distance from the center, \ie, faster and larger in size, than the latter.  
Comparing the pairs, knots `A' and `D' appear to expand at the same speed. 
The hot bubble S$_{\rm A, B}$, observed at high latitude, is not close to the ellipsoidal shell. 
The expansion velocity of these three substructures is faster than the maximum rate of expansion of the EM2 shell
\ie 54 {\kms} at the apex of the main shell.
All five knots are independent structures separated from the main shell, expanding at a very high speed.

We propose that the two pairs of knots `A' -- `D' and `B' -- `C' in approximate point-symmetric positions are virtually aligned on the same axis of expansion, \ie latitudes $\phi \sim \pm 34.9(\pm 0.4)^{\circ}$ and $\phi \sim \pm 34.1(\pm 0.5)^{\circ}$ (whose mean is about $\phi \sim \pm 34.5(\pm 0.6)^{\circ}$). 
Meanwhile, two spots S$_{\rm A}$ and S$_{\rm B}$ of the hot bubble expand at a speed of 132 -- 148 {\kms} at a radius of  16.4 -- 17.2$''$.

\begin{table}[ht]
\caption{ Expansion velocities of two knots and hot bubble points. 
 }

\hspace{-0.7in}
\begin{tabular}
{cccc cccc}
\hline 
Knot    & $d$(ma)$''$ &  V$_{\rm rad}$(obs)   & EM2($r'', \phi_{\beta}^{\circ}$)   & V$_{\rm exp}$(CM3) &  $\Delta r$  &   $|\Delta V_{\rm los}|$    & ($r''$), V$_{\rm exp}$  range \\
\hline
A     & 9.8 & -37$\pm$6  &   (16, 35.3)   &  62.3      &  2$''$(m) vs. $2''$(o)  & 17.4(m) vs. 19(o)  &  {\bf  (16), 62.3  }    \\
      & (11.5 -- 13.3)   &    &      &        &       &        &  { (15--16), 55.1--72.5 } \\
 D   & -10.0 &  ~35$\pm$3  & (16.5, -34.5)  &  57.8    &     1.5$''$(m) vs. $1.5''$(o)  & 11(m) vs. 11(o)  &  {\bf (16.5), 57.8}     \\
    &   (-10.0 -- -11.5)   &    &     &   &     &      &  (16.5--17.5), 57.8--68.8 \\
S$_{\rm A}$   &  13.7 & -14$\pm$3 &  (16.4, 65.6) & 132.2  &    1.0$''$(m) vs. $1.0''$(o) & 7.3(m) vs. 4.5(o)      &   {\bf (16.4), 132.2}   \\ 
     & (13.2 -- 14.2)   &    &     &   &     &      &  {(15.82--16.97), 129.3--136.6}  \\ 
S$_{\rm B}$   & 15.3 & 5$\pm$3  &  (17.2, 73.6) & 147.8  &  1.5$''$(m) vs. $1.5''$(o) & 10.3(m) vs. 4.5(o)      &   {\bf (17.2), 147.8}    \\
     &  (14.8 -- 16.3)   &    &     &   &     &      &  (16.58--18.28) 142.5--152.8 \\ 
 \hline
\end{tabular} \\
\vspace{0.05in}

{Notes. All Velocity units: {\kms}. 
See Table C in Appendix C.  
Column (2): a spatial distance to the projected point on  the major axis, \ie $y_1$ in Appendix C. 
Column (3): The observed radial velocities relative to the major axis correspond to line-of-sight velocities, \ie V$_{\rm rad}$(obs) = V$_{\rm los}$(model).
Column (4): radius and latitude to the knot $\phi_{\beta}$.
Column (5): predicted expansion velocity.
Column (6): $\Delta r$ is the knot size (see Appendix C CM3(m) and Observed Dimensions (o) in Figure 2).
Column (7): the line of sight velocity difference (or range) of the model (m) and the observed knot dimensions (o) in Figure 2. 
Column (8):  summary of knot radius range and expansion velocity range from two lines. }
\label{tbl-4}
\end{table}

\begin{figure*}[ht]
\centering
\includegraphics[width=100mm]{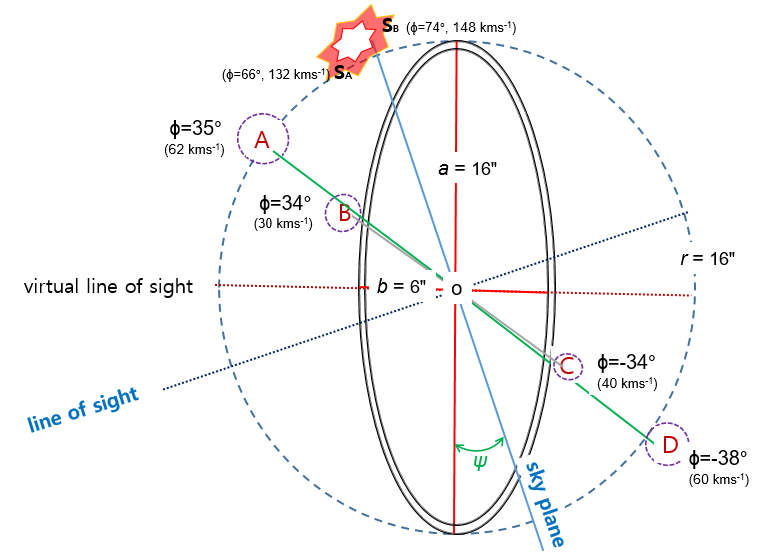}
\caption{Schematic diagram of the main shell,  four knots, and a hot bubble (FLIERS). 
The projection angle of the major axis onto the sky plane is $\psi \simeq$ 18.3$^{\circ}$. The four knots expand along the nearly same axis, $|\phi| \simeq 35(\pm 3)^{\circ}$ relative to major axis. The approximate knot velocity is given in parentheses.
See Tables~\ref{tbl-2} -- \ref{tbl-4} for densities and expansion velocities. 
 }\label{fig3}
\end{figure*}

Figure~\ref{fig3} shows a schematic diagram of an ellipse with $a \sim$ 16$''$  and $b \sim$ 6$''$   representing the ellipsoidal shell, with the surrounding knots and a hot bubble (S$_{\rm A}$ and S$_{\rm B}$). 
The virtual line of sight indicates a line parallel to the $x$-axis or the minor axis, and the line of sight is the radial direction by us.  
We did not mark the outer torus shells at low latitudes \citep{ste09} and the equatorial shell observed in Paper I. The spectral image along the major axis could not reveal the outer shell clearly at the equator due to the relatively strong stellar continuum but did identify its presence and the blue-shifted side low-latitude tori (in Figure 2(e)).

The four knots are detached from the main shell (see PV density diagram in Figure 2).
The positions and latitudes of `B,' `C,' `A,' and `D' imply that they are two pairs, each of them emitted at the same period. Most surprisingly, it looks like all four knots are aligned along nearly the same axis.
 
The knot pair `A' and `D' which are largely different from the prolate shell, also have large relative distances from the CSPN (their latitudes) than the other pair `B' and `C.' 
The latter pair  `B' and `C'  expands at a faster velocity than the shell but looks closer to the primary ellipsoidal shell (see Figure~\ref{fig2}).

The dimensions of knot pair `A' and `D' are more substantial than the other pair `B' and `C' sizes, or the main shell thickness. Assuming the distance to NGC 7009 is 1.15 kpc by \citet{2018kim} and using the average model predicted expansion velocities in Table~\ref{tbl-3}, then `B' and `C' were released from the center 1370$\pm$240 years ago, 
and the rapidly expanding `A' and `D' in Table~\ref{tbl-4}  would have appeared 1480$\pm$80 years ago.
The hot spot was released very recently, about 655$\pm$22 years ago.

Some irregularities and subarc-scale smaller rims in the primary shell could be evidence of the multiple mass-loaded outflows from the CSPN during the post-AGB phase, as seen in M 2-56 \citep{2010San}. 
Or it could be an earlier shell, ejected from the CSPN long before the ejection of the outer spherical shell and the primary prolate shells identified in the equatorial zone.

Assuming that the observed expansion rates of the main and outer shells of the equator of Paper I are the constant expansion ones,
The main and outer shells are 1450$\pm$60 years and 2140$\pm$45 years old, respectively. 
The outer and inner main shells mentioned in Paper I may have formed due to a large amount of gas emitted from the CSPN  with an interval of about 700 years.  Knots `A' and `D' appear to have been released at nearly the same period of the main shell development, followed by the release of  `B' and `C'  about one hundred years later than knots `A' and `D.' 

The two pairs of large knots, found near the polar axis, appear to be formed by entirely different mechanisms. For example, the mass flows out of the companion star, and the hot CSPN thrusts back the bipolar cone shell.
The other corresponding knot `X' of the E-NE apex is less advanced but looks like a hot bubble.
The W-SW hot bubble at the W-SW apex (including the `X' at the E-NE apex) appears to have been released quite recently than all the other structures.

In a binary system, ejections from the common envelope of two stars can occur in the direction of the binary system's orbital axis \citep{2020HS}.
If such an ejection is the cause of the four knots, we reasonably understand why the direction of the major axis of the PN ellipsoid shell is different from the direction of expansion of the four knots. 
NGC 7009 may be a PN, that evolved from a semi-detached binary system.

\section{Conclusions}

We analyzed the 2-D [{\sii}] spectral images secured with the Keck HIRES along the major axis of the nebula. 
The 2-D [{\sii}] image shows the out-most low-excitation zone, \ie the outer boundary of the shell.  
The PV density distribution obtained in the major axis slit entrances shows that the local density of the shell and knots obtained along the line of sight is log~$N_{\rm e}$  $\sim$  {2.8}  --  4.7 dex {\cc}, a much wider range than those integrated over the line of sight, log~$N_{\rm e}$  $\sim$  3.4  --  {3.9} dex {\cc}. 
The derived PV density distribution diagram shows well-separated knot features from the main shell, which are much more dynamic than 2D spectral images or 1D high-dispersion observations (\eg \citealp{ha95a, ha95b}).

Considering that the expansion axis is projected onto the sky plane, it is vital to know the projection angle of the major axis to the sky plane to study the actual expansion speed and its expansion axis of the substructure near the high latitude shell.
We derived a major axis projection of the prolate shell against the sky plane,  $\psi$ $\sim$ 18.3$\pm 2$$^{\circ}$, 
different from other research proposals, \eg  \citet{1985RA}.
The derived projection angle of the major axis of the primary shell will help avoid confusion in future studies of kinematics and nebula shell formation.

The observed spatial and velocity dimension of the knots and the hot bubble in Figure 2 became a constraint in deriving the expansion velocity range and the distance ranges in the model simulation. In Tables~\ref{tbl-3} and~\ref{tbl-4}, we summarized the expansion velocities and radius ranges for four knots and two points in the W-SW hot bubble.

With the help of the ellipsoidal and spherical models, we estimated the radius ranges of the four knots and a hot bubble, their expansion velocity ranges, and the expanding directions (latitudes), respectively. 
The four knots at mid-latitude $\phi = \pm34.5(\pm 0.6)^{\circ}$ expand in symmetrical positions along the same axis.  One pair expands at 34.9($\pm$5.4) {\kms} close to the main ellipsoidal shell, and the other expands rapidly at 60.1($\pm$2.3) {\kms} at a distance of $r \sim 16''$. 
These prominent knots can also be identifiable in the [{\nii}] and [{\oiii}] spectral images by \citet{sab04}. 
In the latitude range $\phi = $65 -- 75$^{\circ}$, the hot bubble of a relatively large structure expands rapidly with a velocity of 130 -- 150 {\kms}.

After forming the main shell, the four knots and hotspots (a hot bubble) started to release from the center, in the order of two knots `B'+`C,' a hot bubble, and the other two knots `A'+`D,' and evolved into the present shape. And four knots, `B,' `C,' and `A,' `D,' are all expanding symmetrically along the same axis. 
These most prominent knots, `A' and `D,' and the hot spot appear to be formed by another formation mechanism rather than a simple stellar wind involving the ellipsoidal main shell formation. An ellipsoidal shell can be formed by a hot CSPN starburst against thermal pulses in an AGB star.

We conclude that there were several distinguished ejections responsible for the knots and sub-arcsecond scale substructures and the multiple component line profiles related to the central star evolution.  
The effective temperature of CSPN is known to be higher in the early stages.
Changes in central star temperature can lead to faster propulsion gases from the CSPN, resulting in increased terminal speed.
\citet{hyu14} discussed the radial change of the expansion velocities of sub-shells, responsible for the various ions along the minor axis. 

Spectral images along the major axis and PV diagrams show the presence of well-separated, low-latitude outer toric shells from the main shell. 
These tori seem related to the outer shell studied in Paper I, but we have not analyzed them in this study. 
Further investigation is needed, along with information about the physical condition available through nebula diagnostics.
These examples are perhaps IC 4997 and Hubble 12, both of which are in the early stages of the second post-AGB, so the dense shell exists closer to the CSPN inside the primary shell
 \citep{1994hyung, 1996hyung, 2000LH}.

\section*{Acknowledgments}
S.H. is grateful to the late Prof. Lawrence H. Aller of UCLA, who conducted the Keck HIRES observation program together.  S.H. also would like to thank F. Sabbadin for encouraging by sending a paper by \citet{sab04} in 2004 and  Prof. Francis P. Keenan at the Queen's University of Belfast for providing additional information on the [{\sii}] atomic data.
We express our gratitude to the anonymous referee who carefully read this paper and made many valuable comments, especially on PyNeb and other atomic constants. 
We thank Dr. Sung, E.-C. at KASI and Dr. Son, D.-H. at SNU for their help with the 2-D spectral data reduction.
S.-J.L. and S.H. would like to acknowledge support from the Basic Science Research Program through the National Research Foundation of Korea (NRF 2020R1I1A3063742; NRF 2017R1D1A3B03029309).
M.O. was supported by JSPS Grants-in-Aid for Scientific Research(C) (JP19K03914).

\vspace{5mm}

\facilities{Keck:10m, HDS}

\software{IRAF (Tody 1986,1993), Starlink \citep{2014star},            Cloudy \citep{2017RM}}

\bibliography{main.bbl}{}
\bibliographystyle{aasjournal}

\newpage

\begingroup
\let\clearpage\relax

\renewcommand{\thefigure}{A}
\setcounter{figure}{0}

\begin{center}
{APPENDIX A. ELLIPSOIDAL EM1 SHELL WITHOUT PROJECTION } \\
\end{center}

Assume a hollow prolate elliptical shell for a general appearance of the inner main shell, seen in Figures 1 and 2 (see also \citealt{ste09, sab04, 2004Fern}). Such an ellipsoid can be related to the expansion velocities, increasing with latitude.

\begin{figure}[ht]
\centering
\includegraphics[width=120mm]{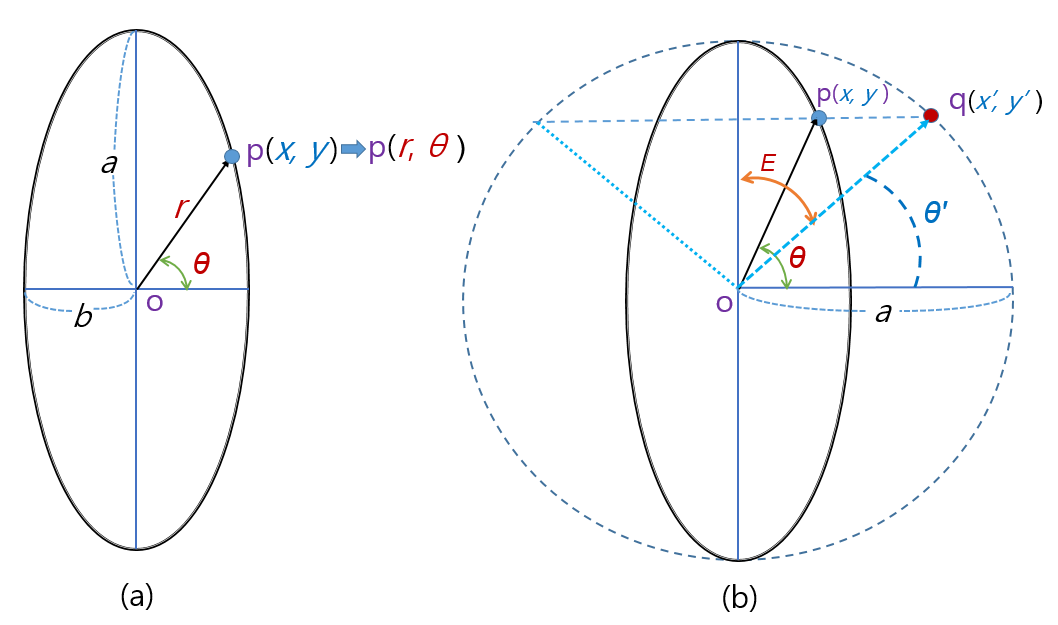}
\caption{Prolate ellipse (EM1).
 }
\label{figA}
\end{figure}

Figure A1 shows the projection of the aforementioned prolate spherical shell (EM1) whose major axis is aligned with the sky plane, \ie an ellipse with eccentricity $e$ = $\sqrt{1-b^2/a^2}$,
$$
\frac{x^2}{b^2} + \frac{y^2}{a^2} = 1~~~  (1)
$$
Figure 2 shows that the semi-major axis radius from the center to the edge of the main shell is 14.5$''$ on the SW side and 15.9$''$ on the NE side, respectively. Hence, we assumed the semi-major axis radius to be  15.2$''$. 
As the semi-minor radius is $b \sim 6''$ (paper I), the axis ratio $a/b$ is assumed to be $\sim$ 2.5. 

Here, $x$ and $y$ are the Cartesian coordinates of a point `p' on EM1.
One can transform the position of a point p($x$,$y$) on the EM1 into spherical coordinate p($r,\theta$) with its radius $r = \sqrt{x^2 + y^2}$  and latitude $\theta$,  
$$ 
\tan(\theta) = \frac{y}{x},~~~~~~ (2)
$$
where $\theta$ is the latitude of the EM1 shell.
The shape of EM1 ($r$, $\theta$) is assumed to originate from a differential expansion: the expansion velocity of the EM1 shell is  proportional to $r$ from the center. 
The expansion velocity at a specific point on EM1, 
$$
{\rm V_{exp}} = 
({{\rm V_{x}}^2 + {\rm V_{y}}^2})^{0.5} = (r/6) \times {\rm V_{e}}~~~~~ (3)
$$
where the expansion velocity at the equator is V$_{\rm e}$ = 21.7 {\kms} at $x$ =  6$''$ and $y$ = 0$''$).
The shell at a specific point p($x,y$) expands with
V$_x$ = 21.7$ \times$ $x/b$ {\kms} (21.7 -- 0 {\kms}) and
V$_y$ = 21.7$ \times$ $y/b$ {\kms} (0 -- 54 {\kms}) for $\theta$ = 0 -- 90$^{\circ}$. For example, we can assume that the vertex point (at the end of the major axis) expands with  V$_{\rm exp}$ $\sim$ 54 {\kms} (or 2.5 $\times$ 21.7  {\kms}).

The above simple case corresponds to the case where the major axis coincides with the sky plane. 
The latitude $\theta$ of a point `p' in EM1 can be obtained  from a different way, \ie using a relationship of an imaginary circle with radius of $a$ and the above mentioned ellipse as given in Figure A(b).  
Figure A(b) shows that $y = a~ {\rm cos}(E)$, $x = b~ {\rm sin}(E),$ and tan($90^{\circ}-\theta$) = $x/y$. Hence, $ 
{\rm tan}(90^{\circ}-\theta) = x/y  = (b/a)~{\rm tan}(E)$, and 
$ {\rm tan}(\theta) =  \frac{(a/b)}{{\rm tan}(E),}$ where the eccentric anomaly $E$, cos($E$) = $y/a$. The spatial distance $y''$ can be obtained from Figures 2(d) --(f). 
This second method helps interpret the observed spectral data as an elliptical or circular shell ($r = a = 16''$), depending on the circumstances. 
When the observed point is `q' on a circular shell, as shown in Figure A(b), $x'$ = $r$~cos($\theta ')$ and $y'$ = $r$~sin($\theta ')$ where latitude $\theta '$ = $90^{\circ} - E$. Note that $\theta '$ in a circular shell is different from $\theta$ in an ellipse.    \\

\setcounter{section}{1}

\renewcommand{\thefigure}{B\arabic{figure}}
\setcounter{figure}{0}

\renewcommand{\thetable}{B}
\setcounter{table}{0}

\begin{center}
APPENDIX B. ELLIPSOIDAL EM2 SHELL AND SKY PLANE 
\end{center}

We compare the expansion velocities for the assumed prolate shell with the actual expansion velocities of the four bright knots and hot bubble structures shown in the PV diagrams of the density distribution in   Figure~\ref{fig2}.
Figure~\ref{fig2} shows an elliptical rim appearance that is not symmetric relative to the vertical line (white). Then we need to find the projection angle of the prolate shell to the sky plane (see Paper I). 
The tilted kinematic baselines (sky blue) in  Figures~\ref{fig2}(d), (e), and (f) represent the correct kinematic center at each specific latitude ($\phi$).
The newly derived kinematic baseline at the vertex (or pole $\phi$ = 90$^{\circ}$) indicates that the major axis of the ellipsoidal shell deviates from the sky plane.

\begin{figure*}[ht] 
\centering
\includegraphics[width=180mm]{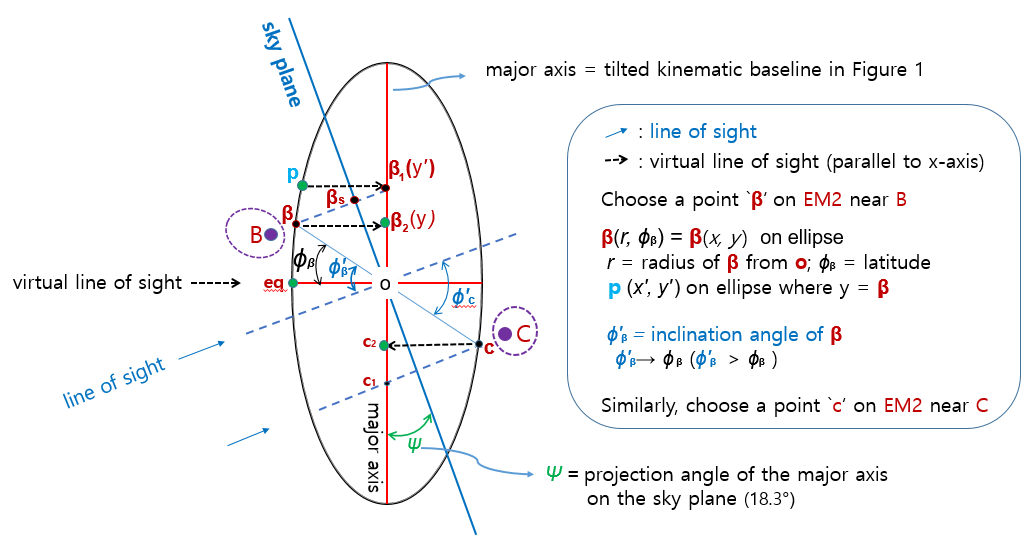}
\caption{Schematic diagram for a prolate ellipse  (EM2) projected to the sky plane with $\psi = 18.3^{\circ}$.
Points p($x', y'$) and $\beta$($x, y$) are on EM2.~  $\overline{{\rm o} - \beta_s}$ = 5.5$''$ (Figure 2);   ~$\overline{{\rm o} - \beta_1}$ = 5.8$''$;  
  ~$\overline{\beta_1 - \beta_2}$ =~$x'\cdot$tan($\psi$) where a point ($x, y$) is on EM2 and $y$ =~$\overline{{\rm o} - \beta_1}$.   $\beta(x, y)$: $y$ = $y'$ - $\overline{\beta_1 - \beta_2}$. See Table~\ref{tbl-b}. }
\label{figB1}
\end{figure*}

The observed radial velocity is 17 {\kms} at the 16$''$ in Figure 2(d), and the expected expansion velocity is 54 {\kms} at the vertices of the prolate ellipsoid. Both represent the relationship of 17 {\kms} = 54  {\kms} $\times$ sin $\psi$. So, the projection angle of the major axis of the prolate shell on the sky plane is $\psi$ = 18.3($\pm$2)$^{\circ}$, where $\psi$ = 90$^{\circ}  - i$ for inclination angle $i$.  The error reflects the uncertainty of the PA choice of the major axis plus about 1-pixel uncertainty in the y-axis of the image in Figure 2. 
This projection angle allows us to find the semi-major radius of the elliptical shell 
as 16$''$ (= 15.2$''$/cos($\psi$)).

Figure B1 shows a schematic diagram of a prolate ellipse (EM2) with an inclination to the sky where $a$ is $16''$ and $b$ is $6''$. The expansion velocity for a point in EM2 can be estimated from coordinates ($x, y$) or ($r$, $\phi$) in Figures~\ref{figB2}(a) and (b). 
Table~\ref{tbl-2} suggests the possibility that knots `B'  and `C'  are relatively close to the EM2 shell, unlike the others.
The schematically drawn elliptical diagram in Figure B1 shows the positions of knots `B' and `C,' the direction of the observed line of sight, and the virtual line of sight by an imaginary observer are marked.

The major axis of EM2 is not aligned along the sky plane, so the latitude $\phi$ of the point $\beta$ near the knot cannot be inferred without taking into account its projection ($\psi$). 
The virtual line of sight is parallel to the x-axis of EM2, so the virtual observer can see the unprojected major axis as shown in Figure A1. 
Figures~\ref{figB2}(b) and (c) briefly describe the current complex situation: the virtual line of sight is parallel to the horizontal axis in Figure~\ref{figB2}(b), but the actual line of sight is inclined relative to the horizontal axis. 

The $\beta_1$ point in Figure~\ref{figB1} is the projection of $\beta$ onto the major or $y$ axis.
The radius of the  semi-major axis is $a$ = 16$''$ considering $\psi =$ 18.3$^{\circ}$. In Figure 2, the point $\beta$ at a distance of 5.8$''$ selected near the knot `B' will be projected on the major axis at $\beta_1$ whose distance from the center is 5.8$''(= 5.5''$/cos({$18.3^\circ$})). See Figure~\ref{figB1} for additional information on other key locations.
Like all points on the EM2 ellipse, the coordinate p($x', y'$) of the point `p' or $\beta(x, y)$ follows the elliptic equation (1) or $x = b~(1-y^2/a^2)^{0.5}$.

Next, we need to derive the expansion rate for $\beta$($x, y$) on an EM2 elliptical shell.
From a rectangular triangle  
$\triangle$($\beta_1$ -- $\beta$ -- $\beta_2$) and $p$ in Figure B1, we find that $y'$ =~$\overline{{\rm o} - \beta_1}$, and ~$\overline{\beta_1 - \beta_2}$ = ~$|x|$tan($\psi$).  
So the $y$ of the $\beta(x, y)$ coordinates of a point on the EM2 ellipse shell is  
$$
y =    y' -   |x|{\cdot}{\rm tan}({\psi}). ~~~~~~ (4)
$$
By applying this relation to equation (1), we get the second derivative equation with respect to $x$: 
$$
(\frac{1}{b^2} + \frac{{\rm tan}^2(\psi)}{a^2} )x^2 
- \frac{2 y' \cdot {\rm tan}(\psi)}{a^2}x + (\frac{y'^2}{a^2}-1) = 0.  ~~~(5) 
$$
For given $a = 16$, $b = 6$, $\psi$ = 18.3$^{\circ}$, and $y = 5.8''$, we find  $\beta$($x, y$) = $\beta$(-$5.8'', 3.9''$) or $\beta$($r, \phi_{\rm \beta}$) = ($7.0''$, 33.6$^{\circ}$)  for  $\beta$ on EM2.  
Figure~\ref{figB2}(c) shows that knot `B' is attached to the EM2 shell and the size of the knot is $\Delta y (= \Delta x) \sim$ 1.5$''$ or $\Delta r \sim 1.5''$. Then the distance to the `B' knot ranges $r_{\rm max} = 8.5'' (= 7.0 + 1.5)$ -- 8.8$'' (= 7.3 + 1.5)$.
Note that the EM2 latitude ($\phi_{\rm p}$) is smaller than the  EM1 latitude ($\theta$) by 11.0$^{\circ}$, \ie $\phi_{\rm p}$ = 33.6$^{\circ}$ (EM2) $<$   $\theta$ = 44.6$^{\circ}$ (EM1) for the point nearby Knot B. 
The correct latitude is a prerequisite correctly to understand the relationship between the expansion velocity of the shell and its decomposition component 
along the line of sight.  
Table~\ref{tbl-b} lists the aforementioned coordinates for some specific points on the EM2 ellipse, assuming that these are close to four knots and hot spot S$_{\rm A}$.

\begin{table}
\caption{ Expansion velocity derivation with EM2. 
 }

\hspace{-1.in}
\begin{tabular}{lccccc ccccc}
\hline 
Knot & $\overline{{\rm o}-\beta_s}$  & $\beta_1(y')$  & $\beta$($x, y$) & $\beta$($r, \phi_{\rm p}$) [$\phi'_{\rm p}$] & V$_{\rm exp}$[$r''$] &  (V$_{\rm los}$, V$_{\rm sky}$) &  V$_{\rm rad}$(obs)  & $|\Delta$ V$_{\rm los}|$  & V$_{\rm exp}$(knot) & $r''_{\rm max}$   \\
  & ($''$)      &   ($''$)  &    ($~'', ~''$)  &   (~$'', ~^{\circ}$) [$^{\circ}$]  &  EM2 &  EM2  &   &   & (estimated) &  [V$_{\rm exp}$]  \\
\hline
B  &   5.5  &  5.8   &  (-5.8, 3.9)  & ({\bf 7.0}, 33.6) [51.9] & {\bf 25.3 [7.0]} & (-15.6, -19.9) & -18$\pm$3  & {\bf 2.4 } & {\bf 29.2 }   & 8.5 -- 8.8  \\
  &   6.0  &  6.3    &  (-5.8, 4.4)  & (7.3, 37.3) [55.6] & {26.2 [7.3]} & (-14.8, -21.6) & -18$\pm$3  & { 3.2}  & {31.9 }    & [29.2--31.9] \\
C  &   -5.5   &  -5.8  &    (5.8, -3.9)  & ({\bf 7.0}, -33.6) [-51.9] & {\bf 25.3 [7.0]} & (15.6, 19.9) & 25$\pm$5 & {\bf 9.4} & {\bf 40.5}   &  8.3  -- 8.5 \\
  &    -5.0  &  -5.3  &    (5.9, -3.4)  & (6.8, -29.8) [-48.1] & {24.4 [6.8]} & (16.3, 18.2) & 25$\pm$5 & { 8.7} & { 37.4}   &  [37.4--40.5] \\
A  &   9.3  &  9.8  &   (-5.2, 8.1)  & (9.6, 57.4) [75.7] & {\bf 34.7} [9.6] & (-8.6, -33.7)  & -37$\pm$6 & 28.4 & ~~150 (x)  &   (x)  \\
D  &   -10.3   &  -10.9  &    (4.9, -9.3)  & (10.5, -62.2) [-80.5] & 37.9 [10.5] & (6.2, 37.4) & 35$\pm$3 & 28.8 & ~~213 (x)   & (x)  \\
S$_{\rm A}$ &   13 &  13.7  &    (-3.8, 12.5)  & (13.0, 73.2) [91.5] & {47.1 [13.0]} & (-1.2, 47.0) & -14($\pm$3) & 12.8 & ~~545 (x)  &  (x) \\
 \hline
\end{tabular} \\
\vspace{0.02in}
{ All Velocity units: {\kms}. Column (2): the radial distance from the center. 
Column (3): the projected spatial distance along the major axis. Columns (4) -- (5): positions on EM2, close to the knots.  $\beta$($x, y$) and $\beta$($r, \phi_{\rm p}$) for knot position `p,' \eg $\beta$($r, \phi_\beta$), which can be applied to knot `B,' to `C,' etc. The inclination angle of each knot is $\phi'_{\rm p}$, \eg $\phi_\beta'$ for $\beta$ (see Figure B1).
Column (6): the expansion velocities of the EM2 shell using equation (3).  
Column (7): the decomposition velocities (the line of sight and the sky plane components) of the EM2 shell. 
Column (8): observed radial velocity along the line of sight, relative to the new kinematic baseline in Figure~\ref{fig2} (or the major axis in Figure~\ref{figB1}). Column (9): $|$V$_{\rm rad}$(obs) - V$_{\rm los}|$. 
Column (10): estimated best expansion velocities for knots: V$_{\rm exp}$(knot) =  V$_{\rm rad}$(obs)/cos($\phi'_{\rm p}$). 
 {(x)}: (not adopted).
Column (11): Best estimation of radius and velocity range. 
 The distance from the center to `B' and `C' is estimated considering the size of the knot. The velocity range of `B' and `C' in the last column is adopted in column (10).
When the derivation is unreasonable, it is indicated by (x). See the text. 
\\
}
\label{tbl-b}
\end{table}

\begin{figure}[ht] 
\centering

\begin{minipage}{0.45\textwidth}
        \centering
        \includegraphics[width=1.25\textwidth]{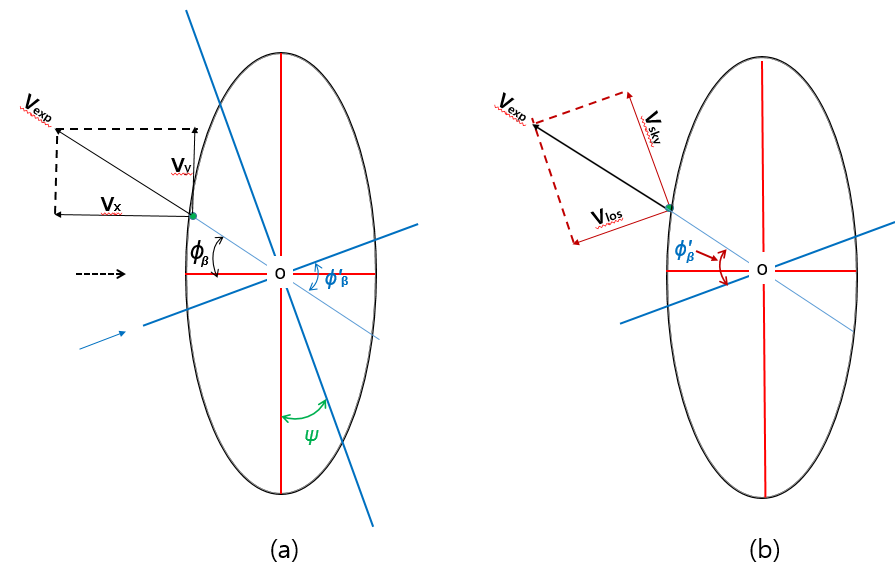} 
\end{minipage}\hfill
    \begin{minipage}{0.5\textwidth}
        \centering
        \includegraphics[width=0.47\textwidth]{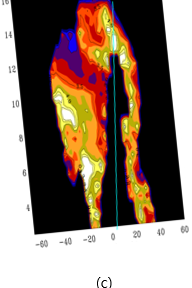} 
    \end{minipage}
\caption{Expansion velocity and the line of sight component. (a) Expansion velocity vs. latitude (V$_{\rm exp}$ and $\phi_\beta$). (b) Expansion velocity vs. radial velocity along the line of sight (V$_{\rm los}$ and $\phi_\beta'$). 
(c) Figure 2(d) was rotated and reproduced relative to the new kinematic center line or the major axis. }
\label{figB2}
\end{figure}

Figure B2(a) shows a relationship between the expansion velocity V$_{\rm exp}$  and its components (V$_{x}$, V$_{y}$), which depends on $\phi_{\beta}$, while Figure B2(b) shows different viewpoint velocity components (V$_{\rm los}$, V$_{\rm sky}$) along the line of sight and the sky plane tangential to the line of sight, respectively.  
Figures B2(a) and (b) illustrate the latitude angle ($\phi_{\beta}$) from the equatorial plane and an inclination of the expansion axis relative to the line of sight ${\phi_{\beta}'}$ (${\phi_{\beta}} <  {\phi_{\beta}'}$) for $\beta$. The angle of inclination is essential information to compare the EM2  prediction with the observed radial velocity of the knot. 
As mentioned, the rate of expansion of the EM1 shell is proportional to its distance (radius) from the center.  
But one did not observe the expansion velocity of knots directly, so we must find the line of sight component after finding the expansion velocity from the EM2 shell.

Figure B2(a) shows the relation of the expansion velocity of a point  $\beta$($r, \phi_{\beta}$) on  EM2 shell (here  $\phi_{\beta}$  is similar to $\theta$ in Figure A).  
Knowing the coordinates and velocities expanding radially outward from the center, we can find the velocity components according to a particular point of view in the model geometry and compare them to those observed in Figure B2(c).

From equation (3), V$_{\rm exp}$ = 21.7 $\times$  ($r$/6)= 25.3 {\kms} for $\beta$. One can further derive the projected velocity component along the line of sight (see  $\phi_{\beta}'$ in Figure B2(b)):  V$_{\rm los}$ = $-$15.6 {\kms}, which is fairly different from the observed value of   V$_{\rm rad}$(obs) = $-18\pm 3${\kms}. This suggests  V$_{\rm exp}$(B) $\sim$ 29.2 {\kms}.
In a similar way, one can find the velocity components  and compare the model prediction with the observed radial velocity for the knot `C':
V$_{\rm exp}$ = 25.3 {\kms} and  V$_{\rm los}$ = 15.6 {\kms} by EM2  shell, implying V$_{\rm exp}$(C) $\sim$ 40.5 {\kms} from  V$_{\rm rad}$(obs) = $25\pm 5${\kms}. 
Thus, 
the two knots, `B' and `C,' are expanding about 5--15 {\kms} faster than the shell at about 1--1.5$''$ farther than the EM2 shell distance.

Although there is a relatively significant difference, the `B' knot and the `C' knot seem to exist reasonably close to the shell.
Therefore, the expansion velocity of the EM2 shell becomes the basis for estimating the kinematic properties of these knots.
The last column of Table~\ref{tbl-b} provides a summary of the allowable range for the radius and velocity of the knot obtained based on the observed range of knot velocity and size.
On the other hand, hot bubble spots and other knots, `A' and `D' are difficult to analyze their kinematic properties with the help of the prolate EM2 shell, so we conclude that they are structures independent of the ellipsoidal main shell.  \\

\renewcommand{\thefigure}{C\arabic{figure}}
\setcounter{figure}{0}

\renewcommand{\thetable}{C}
\setcounter{table}{0}

\begin{center}
APPENDIX C. SPHERICAL CM3 SHELL  
\end{center}

Table~\ref{tbl-b} in Appendix B shows that the radial velocities of the two knots `A' and `D' cannot be explained with an ellipsoidal shell geometry as in Figure~\ref{figB1}. We, therefore, hypothesize that these knots extend radially outwardly at a relatively high expansion rate independent of the main shell.
In Figure C1, an imaginary circular shell is drawn and the relative positions of the two knots are marked. 
The major and minor axes introduced by EM2 in Figure B1 which are independent of the circle geometry, still retain here. This baseline is useful for displaying the projection angle $\psi$ relative to the sky plane. 
Based on the spatial distance from the center of Figure 2, the relative positions of the substructures were established along with an imaginary circle of $r$.

Figure~\ref{figC1} shows the position, latitude ($\phi_{\rm A}$), inclination angle ($\phi'_{\rm A}$) relative to the line of sight, and coordinates ($x, y$) of knot `A' (and `D') that appear when the observer observes knot `A' located in the virtual circle.
Table~\ref{tbl-c} shows all the step-by-step processes to derive the expansion velocity and the distance range.
We kept the line of sight and the virtual line of sight in the circle diagram in addition to the major and minor axis of the EM2 in Figure B1. The virtual line of sight (parallel to the x-axis) yields the latitude $\phi$, while the line of sight yields the projection angle $\phi '$, in a similar way as in Figure B1. 

\begin{figure*}[ht] 
\centering
\includegraphics[width=180mm]{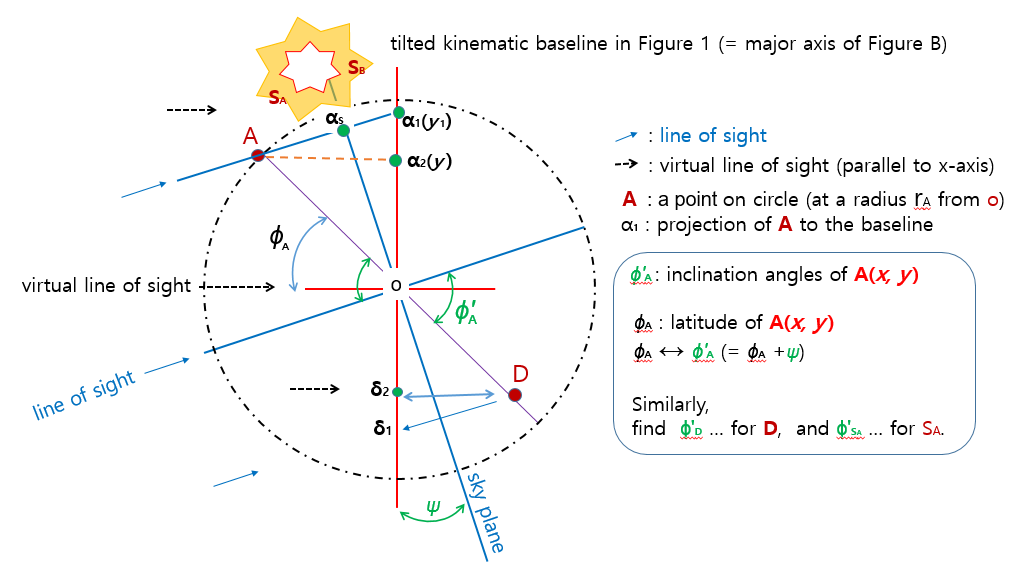}
\caption{Schematic diagram for a circular shell (CM3) projected to the sky plane with $\psi = 18.3^{\circ}$.  The diagram shows that the radius $r$ of point `A' is greater than the distance from the center to the projected point $\alpha_1$ along the Figure B's major axis (the  tilted kinematic baseline in Figure 2): $r_{\rm A} > $ $\overline{{\rm o} - \alpha_1}$ $> \overline{{\rm o} - \alpha_2}$.  
A($x, y$): the coordinates of A on an imaginary CM3 circle with radius $r_{\rm A}$, where $y = y_1 - x\cdot$tan({$\phi_{\rm A}$}) and A($x, y$)  =  
 A($r_{\rm A}, \phi_{\rm A}$). 
 }
\label{figC1}
\end{figure*}

The knot `A' is at a spatial distance of  9.3$''$  from the center (see Figure 2 and Table~\ref{tbl-b}) which implies $\overline{{\rm o} - \alpha_s}$ = 9.8$''$ in Figure C1. Therefore, we know that the knot `A' projects at a point $\alpha_1$ along the tilted kinematic baseline (or the major axis)   at 9.8$''$ from the center. Now introduce an imaginary circle of a fixed radius, \eg  $r = 16'' $ for knots `A' and `D.'

The rectangular triangle $\triangle$(A-$\alpha_1 $-$ \alpha_2$) in Figure C1 gives  $\overline{\alpha_1 - \alpha_2}$   =  $x\cdot$tan($\psi$), so 
$$
y = y_1  - x\cdot{\rm tan}({\psi}), ~~~~ (6)  
$$
where $y_1$ is the distance to $\alpha_1$ from the center along the kinematic baseline, which differs depending on whether the position of knot `A' along the line of sight.

For point `A' located in the line of sight,
$x$-value can be found in the condition of $x^2 + y^2 $ = $r^2$ in CM3 shell, 
$$
(1+{\rm tan}^2({\psi}))x^2 - 2 y_1 {\rm tan}({\psi})
x + (y^2_1 - r^2) = 0  ~~~(7)
$$
which depends on $r$-value.

We need to estimate the expansion velocity and the radius r of point `A' and the direction of the expansion in acceleration proportional to the radius r. Figure C1 shows that point `A' must be in the line of sight between $\alpha_1$ and the observer. Figure C1 suggests that the radius from the center to point `A' is greater than the $y_1$ but likely less than the major axis radius ($a$).
Given the radius and expansion velocity of point `A' on a CM3 circle, the component velocity would be determined as a function of latitude. 

The observed radial velocity is dependent on the knot position along the same line parallel to the line of sight.
The exact location of knot `A' is not immediately known, but it should be located somewhere in the line of sight where it should intersect the virtual CM3 circle, as shown in Figure C2(a).
Based on this location, the latitude $\phi_{\rm A}$ and the angle of inclination $\phi'_{\rm A}$ can be found accordingly (\eg see Figure C1).

 At first glance, there seems to be no way to find the radius $r$ value.
However, there is a way to constrain the constraint to a location, \ie a known knot dimension.
When choosing $r$ for a knot, one should consider the range of expected expansion velocity to fit the observed values.
In Figure 2, the dimensions of knots `A' and `D' are 18 -- 20 {\kms} $\times$ 2$''$ and 10 -- 12 {\kms } $\times$ 1.5$''$, respectively, and the other two `B' and `C'  are 14 -- 15 {\kms} $\times$ 1.5$''$ and 10 -- 12 {\kms } $\times$ 1.5$''$, respectively. 
Table~C1 shows that for the adopted radius $r_1$, the expansion velocity can be derived accordingly.
If this selected radius is not correct, the velocity obtained from another radius $r_2$ at a reasonably close distance will not provide an adequate velocity range.

Positions of two knots `A'  and `D' are projected onto the two points $\alpha_1$ and $\delta_1$, \ie the $y$ values onto the tilted kinematic baseline in Figure 2 (or the major axis in Figure C1). 
Like Figure~\ref{figB1} and Table~\ref{tbl-b}, we find that $\overline{o-\alpha_s}$ = 9.3$''$ and $\overline{o-\alpha_1}$ = 9.8$''$ for knot `A' in Figure~\ref{figC1}. The former `A' has a radial velocity of -37 {\kms} relative to the sky plane, and the latter `D' has  +35 {\kms} in Figure~\ref{fig2}.

As in Figure~\ref{figC1}, the radius of a knot must be greater than the distance from the center to a projected point along the major axis. 
Firstly, we begin our investigation with an assumption that r is equal to $r = a$. 
Figure~\ref{figC2}(a) shows possible positions on a line-of-sight line with different radii for knot A. 
We first compute the case where $r  = 16''$ as in columns (4) -- (6) and then do the same procedure with other radii as in columns (7). 
Column (8) is the derived expansion velocity at the smallest possible radius position of the knot, taking into account the observed blue- (or red-)shifted radial velocity.
As shown in Figure~\ref{figC1}, the expansion velocity can be derived from selected points along the line of sight, while the observed location of $\alpha$ and its projection $\alpha_1$ on the major axis becomes a constraint. Therefore, the range of the expansion velocity and radius to the knot that matches the observed radial velocity and size of the knot is obtained.
The expansion velocity of the knot `A' on CM3 has a relationship,   
$$
{\rm  V}_{\rm exp} = ({\rm V}^2_x + {\rm V}^2_y)^{0.5}  
= ({\rm V}^2_{\rm los} + {\rm V}^2_{\rm sky})^{0.5}   
    = {\rm V}_{\rm los} / {\rm cos}({\phi '}),  
$$  
which should be checked against the observed radial velocities.

\begin{table}
\caption{Expansion velocity derivation with CM3.  }

\hspace{-.8in}
\begin{tabular}{lcccccccc}
\hline 
Knot &  $\overline{{\rm o}  - \alpha_1}$  & V$_{\rm rad}$(obs)  & $\Delta$V$_{\rm obs}$    & CM3($x, y$)  & CM3($r, \phi$) & V$_{\rm exp}$ [$\phi'$]       &  V$_{\rm los}$(CM3) &   $r''$  \\
  & ($''$)      &    &  &  ($'', ''$)  &    ($'', ^{\circ}$)  &  ~~~~ [$^{\circ}$]  &     &  [V$_{\rm exp}$] \\
\hline
A  &  {\bf  9.8}  & -37$\pm$6   &  18--20 &   (-9.4, 6.7)  &  (16, 35.3)  &  62.3 [53.6] &   -37 & 15--16 \\
   & 8.5   & &   &  (-9.5, 5.4)  &  ({\bf 15, 29.5})  &    {\bf 55.1} [47.8]  &    &          [55.1--72.5]  \\
  &   10.5   &  &  &  (-8.8, 7.6)  &  ({\bf 16, 41.0})  &   {\bf 72.5} [59.3] &   &   {\bf  }   \\
D  &   {\bf  -10.9}   &  35$\pm$3 &  10--12  &    (9.0, -7.9)  &  (16.5, -41.6)   & 69.7 [59.9] & 35   &   16.5--17.5  \\
  & {-10.0 }   & &   & (9.8, -6.7)  &    ({\bf 16.5}, -34.5)  & {\bf 57.8} [52.8]  &   &    [57.8--68.8]  \\
   &  { -11.5}   & &  &  (9.6, -8.3)  &    ({\bf 17.5}, -41.1)  &  {\bf 68.8} [59.4]  &    & {\bf }  \\
S$_{\rm A}$  & {\bf 13.7}    & -14$\pm$3  & 4--5  &  (-5.4, 11.9)  &  (16.4, 65.6)      & 132.2 [83.9] & -14  &   15.8--17.0   \\
 & 13.2 & & & (-5.2, 11.5) & ({\bf 15.82}, 65.5) & {\bf 129.3} [83.8]  &  &  [129.3--136.6] \\
 & 14.2 & & & (-5.6, 12.4) & ({\bf 16.97}, 65.8) & {\bf 136.6} [84.1] &    & {\bf }  \\
S$_{\rm B}$  & {\bf 15.3}     &   5$\pm$3 & 4--5 &  (4.1, 13.9)  &  (17.15, 73.6)  &  147.8 [91.9] &      5 & 16.6--18.3    \\
 & 14.8 & & & (3.9, 13.5) & ({\bf 16.58}, 73.7) & {\bf 142.5} [92.0]   &   &  [142.5--152.8]  \\
 & 16.3 & & & (4.4, 14.9) & ({\bf 18.28}, 73.6) & {\bf 152.8} [91.9]    &   &    \\
 \hline
\end{tabular} \\
\vspace{0.05in}

{Notes. All Velocity units: {\kms}. Column (2): the distance from the center along the major axis.  
*: a slightly large $y_1 = 17.2''$ instead of $y_1 \sim 17''$. 
Column (3): observed radial or line-of-sight velocity relative to the new kinematic center line (or the major axis). 
Column (4):   $\Delta$ V$_{\rm obs}$ in Figure 2.
Columns (5) and (6): rectangular and spherical coordinates.
Column (7): predicted expansion velocity at $r''$ and the projection angle of the expansion axis of the knot or the hot bubble relative to the sky plane (see Figure C1). 
Column (8): predicted  line-of-sight (or radial) velocity V$_{\rm rad}$(los). This condition helps determine the projection angle of the expansion axis of the knot and hot bubble. See Figures B2(c) and C1 for  S$_{\rm A}$ and  S$_{\rm B}$. 
Column (9):   Range of expected radially outward distances and expansion velocities. Note that the distance $r_{\rm A}$ from the center to S$_{\rm A}$ is close to the semi-major $a = 16''$, but the distance $r_{\rm B}$ from the center to S$_{\rm B}$ is greater than $a = 16''$. } 
\label{tbl-c}
\end{table}

\begin{figure} 
\centering
\begin{minipage}{0.45\textwidth}
        \centering
        \includegraphics[width=1.25\textwidth]{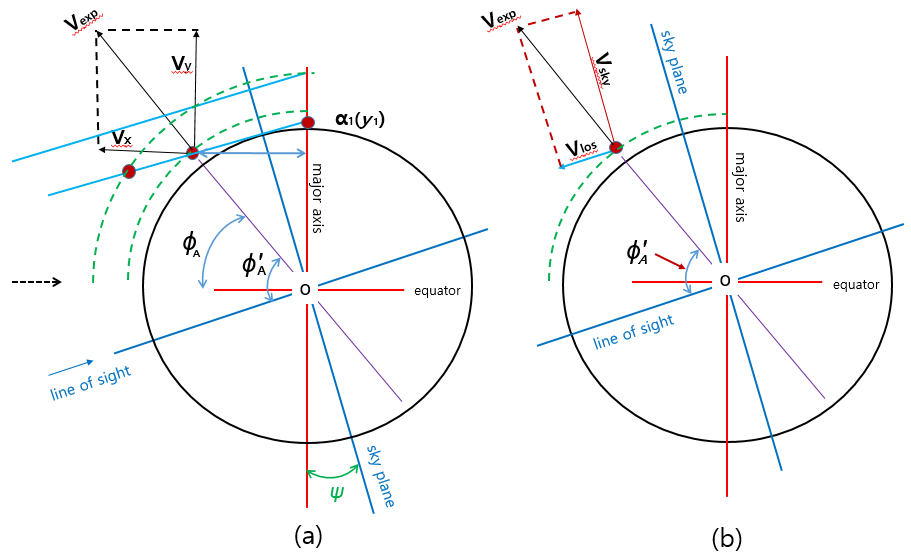} 
    \end{minipage}\hfill
    \begin{minipage}{0.45\textwidth}
        \centering
        \includegraphics[width=0.8\textwidth]{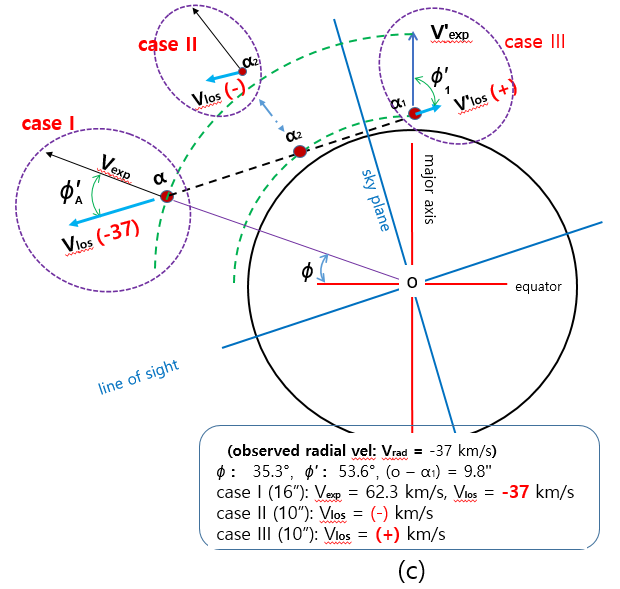} 
    \end{minipage}
\caption{Expansion velocity in a spherical shell.  (a) Expansion velocity vs. latitude ($\phi$, or $90^{\circ} - E$). (b)  Expansion velocity vs. radial velocity along the line of sight ($\phi_A'$). A solid circle: an imaginary circle drawn by a medium value of the knots `A' and `D' distance from the center (dashed arc). (c) The projections of two points $\alpha$ and $\alpha_1$ on the sky plane and their line of sight components. 
Two imaginary circles show that A's radius ($r$) depends on A's position along the line of sight. Thus, the radius $r$ of A determines the latitude of A and the value of the expansion velocity V$_{\rm exp}$. The CM3's predicted line-of-sight velocity for $r$ should match the observed (negative) radial velocity. Improper selection of $r$ can cause the computational model to yield positive line-of-sight values for A. Therefore, the observed radial velocity becomes a constraint. See the text. 
 }
\label{figC2}
\end{figure}

In Figure C2(c), we choose three points $\alpha$, $\alpha_2$, and $\alpha_1$ arbitrarily on a straight line that coincides with the direction of the observer's line of sight. Then the knot's expansion velocities and V$_{\rm los}$ components are illustrated at each point.
Case I shows where V$_{\rm los}$ can induce -37 {\kms} for an appropriately large radius, while case II and case III show a smaller radius close to the main shell boundary.  
Small radius case II requires a substantially large expansion velocity to account for the observed radial velocity.
On the other hand, case III predicts the positive V$_{\rm los}$, which cannot accommodate the observed blue-shift radial velocity.

Three cases are given for the distance from the center to $\alpha_1$, \ie. $y_1 = 9.8'', 8.5''$ and $10.5''$ in Table~\ref{tbl-c}.  Of these, $y_1$ = 8.5$''$ and $10.5''$ give the best expansion velocity range of V$_{\rm exp}$ = 55.1 -- 72.5 {\kms}.
At last, we find a conclusion for radii $r$ and $\phi$s that are consistent with the observed V$_{\rm los}$ = $-37\pm6$ {\kms}  (see Figure~\ref{figC2}(c) diagram): 
(15$'', 29.5^{\circ}$) and (16$'', 41.0^\circ$). 
The radii of knots `A' and `D' are most likely in the range of $r$ = 15
-- 16$''$ and 16 -- 17.5$''$, respectively. Their representative (or average)  latitudes are  $\phi$ $\sim$ 35 and -38$^\circ$, respectively.
We also list the results for knots `A' and `D,' and two hot bubble points S$_{\rm A}$ and S$_{\rm B}$ in Table~\ref{tbl-c}.

\end{document}